\documentclass{article}
\usepackage[utf8]{inputenc}
\usepackage[english]{babel}
\usepackage{epsfig}
\usepackage{graphicx}
\usepackage{listings}
\usepackage{amsthm}
\usepackage{geometry}[margin=1 in]

\usepackage[
citestyle=numeric,
maxbibnames=19,
maxnames=19,
maxcitenames=10,
]{biblatex}
 
\title{Rivendell: Project-Based Academic Search Engine}

\author{Teddy Lazebnik, Hanna Weitman, Yoav Goldberg, Gal A. Kaminka\\ Computer Science Department\\ Bar Ilan University, Israel}
\providecommand{\keywords}[1]{\textbf{\textit{Keywords:}} #1}

\date{ }

\begin{document}

\maketitle

\begin{abstract}
Finding relevant research literature in online databases is a familiar challenge to all researchers.
General search approaches trying to tackle this challenge fall into two groups: \textit{one-time search} and \textit{life-time search}. We observe that both approaches ignore unique attributes of the research domain and are affected by concept drift. We posit that in searching for research papers, a combination of a life-time search engine with an explicitly-provided context (project) provides a solution to the concept drift problem. We developed and deployed a project-based meta-search engine for research papers called Rivendell. Using Rivendell, we conducted experiments with 199 subjects, comparing \textit{project-based} search performance to \textit{one-time} and \textit{life-time} search engines, revealing an improvement of up to 12.8 percent in project-based search compared to life-time search.
\end{abstract}


\sloppy
\section{Introduction}

Finding relevant research literature in a very large online databases is a well-known challenge \cite{JanBrophy,JoeranBeel,Speretta,AndreasHolzinger,Marcelo,Gusenbauer, SteveLawrence, FeiWu}. 
PubMed, arXiv, and other systems hold more than 50 million research papers on the world wide web as of 2010~\cite{ArifE}. Improving research literature search engines is an active field of study. There are two general approaches to search such online databases: \textit{one-time search} and \textit{life-time search}. One-time search engines treat each query independently. Life-time search engines consider data about the user from the user's first search up to the present, to improve their results~\cite{Liu,Speretta,Khopkar,gomez_hunt_2016,Jacobson,Davidson}.

We observe that both approaches ignore unique attributes of the specific task of searching for relevant research. Here, users consider multiple research topics, with multiple queries in each. One-time search answers one query at a time, but does not exploit results of previous queries from the same user. It may exploit previous queries of other users (and from the same user without explicitly be aware to the fact this is the same user) to globally improve results \cite{JoeranBeel}. Life-time search can exploit these results, but, in general, suffers from concept drift, as users switch between research topics intermittently \cite{AztiriaA, Liu, JoaoGama}. 

We posit that in searching for research papers, a combination of life-time search engine with an explicitly-provided context provides a partial solution to the concept drift problem. Specifically, the explicit separation of queries into groups according to different research topics, addresses the problem of local changes in user's searching subject by avoiding studying queries via historical order. For example, user that searches for classical music songs, once looking for pop-song and then search for more classical music. It reduces the noise effects of these changes and improves the overall relevance of results. We call this \textit{project-based search}.

We developed and deployed a project-based meta-search engine for research papers called Rivendell. It improves the relevance of the results provided to the user, over time, by taking advantage of explicitly provided knowledge by the user. Before searching, the user explicitly declares a context (project) for future queries. Then, an ongoing cycling between getting queries from the user (search) and suggesting the next query improves the results.

Using Rivendell, we conducted experiments with 199 subjects, comparing \textit{project-based} search performance to \textit{one-time} and \textit{life-time} search engines. As Rivendell is a meta-search engine, we wished to understand the changes in performance of the one-time search in Rivendell in contrast with arXiv and PubMed, which Rivendell is based on, to get a baseline for further tests. Then, given a personalization algorithm based on natural language processing methods we compared project based search with life-time search. Finally, we investigated reducing the number of needed queries to reach a given precision of relevant results, by suggesting future queries to the user according to the explicit labeling of previous results as relevant or irrelevant.

We conducted three experiments to answer the following questions:
\begin{itemize}
    \item  Does Rivendell provide results in a similar precision of relevance as the leading search engines it is based on?
    \item  Does an explicit separation of the user’s activity to projects for the personalization algorithm improve the precision of relevant results compared to one-time search and life-time search?
    \item Is a suggestion on adding or removing a term from the last given query according to labelling of previous results as relevant or irrelevant improves the search process?
\end{itemize}

The results of the experiments reveal that Rivendell provides similar precision of relevant results as arXiv and PubMed, using only the meta-search engine without personalization or project-based search. Moreover, we found a significant improvement of up to 12.8\% in project based search compared to life-time search, after just six queries. In addition, Rivendell's query suggestion assists the user, and cuts the number of queries needed by half. Therefore, the project-based search approach helps information systems to leverage explicit tagging by expert users to improve the precision of the results obtained in quires over time. 

\section{Motivation and Related work}

Finding papers relevant to one's research is an every-day challenge for scientists around the globe~\cite{JanBrophy,JoeranBeel,Speretta,AndreasHolzinger,Marcelo,SteveLawrence}. 
Multiple technology approaches for searching have been offered, falling into two groups: \textit{one-time search} (each query treated independently), and \textit{life-time search} (results depend on previous queries). 

One-time search engines, treat each query independently, not using the user's search history as a result, one-time search does not exploit the full potential of search engines in the manner of improving the precision of relevant results over time \cite{Chuklin}. 

Life-time search increases the query's precision over time, but suffers from \textit{concept drift}~\cite{AztiriaA, KuhA, RicciF, JoaoGama, atypical_search}. In information systems that use historical data about a user are sensitive to local changes and quick global shifting of the user’s search intent~\cite{Speretta,BernardJ,Liu}. This sensitivity causes the system's results to degrade in quality with time. 

The user's historical segmented data allows the study of the representative concepts of the user~\cite{JaimeTeevan}.
One method is searching on meta-data rather than the data itself \cite{Marcelo}. This approach is not adjusted for search engines but more suitable for recommendation systems.
Some methods of handling recommendation and searching build a profile of the user and filter or reorder results according to each profile \cite{BernardJ}. This methodology uses the life-time data the system gathers.

One approach is to improve the search results using additional data about the user. For example, Hersh et al.~\cite{WilliamHersh} utilize implicit user data to improve results, while \cite{Jacobson} uses meta-data from other sources of information about the user. Other search engines take advantage of both explicit and implicit data (hybrid) from the user~\cite{gomez_hunt_2016, query_context_1}.
For example, Netflix's search engine operates in two different ways. In case the query is related to videos or meta-data about the videos (e.g., actors) in Netfilx's corpus, the algorithm uses explicit data. Alternatively, where there is no such data, the algorithm treats the problem as a recommendation problem taking into consideration the user's implicit data in the form of historical views and classified test. Both approaches lack the ability to handle local changes in users’ behavior and implicit change over time. In comparison, Rivendell uses \emph{explicitly} provided data to overcome concept drift.

Some search engines try to handle concept-drift that originated in taking into consideration long historical activity by taking into consideration only a short-term browsing context \cite{search_no_long_tail}. This approach assumes that in a single session (or any implicit divider), a user has only one intent. In the scope of an academic paper's search, such an assumption is problematic due to the unclear nature of the researching process, where the user does not necessarily have one explicit intent.  

An approach that shown to well handle the concept-drift by detecting atypical search sessions using historical data of the user proposed by Eickhoff et al.~\cite{atypical_search}. This approach personally improves the results for each user. Nevertheless, it assumes that a relatively long (several dozens of searches) historical data of the user is available. Such an assumption is problematic for new (cold-start) users as they do not experience the improvement over a long period of time, while Rivendell performs online learning for each query which results in changes in the query's precision relatively fast. 

Some meta search engines use ensemble learning between several searching methodologies with different ``relevance'' metrics (e.g., relevance ranking functions) to improve the relevancy of the results~\cite{RManmatha}. Rivendell takes advantage of this approach by using PubMed's and arXiv's search engines and adding an additional ``relevance" metric based on the context (project) of the query.

Table (1) presents a few popular searches and recommendation systems divided into two features: one-time or life-time search and the sort of data a user provides explicit, implicit, or combinations of them (hybrid). It is possible to see that one-time search engines cannot use only implicit information on the user as they facing cold start conditions \cite{RicciF} and therefore are not able to personalize relevant results.

The life-time search engines are facing a challenge named \textit{concept drift} accruing as users change there objective over time in unforeseen ways while the system performing a learning algorithm on their behavior. The example of Jane at the beginning of this section demonstrates the challenge of learning the users' objectives as one changes the objectives rapidly. 

Gama et al. \cite{JoaoGama} present a survey of techniques and algorithms addressing concept drift in online supervised learning scenario when the relation between the input data and the cost function changes while learning. In the case of searching by researchers, the challenge of concept drift is increased as a result of the unclear nature of the research and discovery process. Indeed, as a special type of searching, dedicated solutions have been developed and studied to improve the performance of academic search engines \cite{Gusenbauer,JoeranBeel,JanBrophy}; contrary to these solutions, Rivendell asks the user to provide an explicit meta-data about the searches and does not try to find it by itself.

In addition, Rivendell provides a next query recommendation algorithm. The next query recommendation aims to increase the query's precision and by that to help users to complete their intent. Corpus of query's and their results (search logs) can be used to train deep learning models to have a good understanding of complex queries and to suggest next queries \cite{next_query_1, next_query_2}. Specifically, using a pre-trained BERT model \cite{bert} on such corpus shows to well perform in the next query suggestion task \cite{next_query_1}. Nevertheless, this approach is not personalized per user and highly dependent on the relevancy of the train data (e.g., the search logs). 

One approach takes into consideration the context of the user's queries which is represented by a short sequence of queries issued by the same user immediately before the query the algorithm needs to generate \cite{next_query_2}. This is similar to the tail approach used in life-time base search engine and suffers from concept-drift as well. Rivendell asks the user to provide an explicit context for the quires, and by that avoid finding the context from implicit data.

\begin{table}
\begin{center}
\begin{tabular}[c]{|l|l|l|}
         \hline
     &One-time&Life-time\\
         \hline
Explicit & \vtop{\hbox{\strut WhatShouldIReadNext search}\hbox{\strut CiteSeer}\hbox{\strut Manmatha at al. 2001 \cite{manmatha_2001}}}
&  \vtop{\hbox{\strut Qoura question's }\hbox{\strut recommendation}}\\[6ex] \hline

Implicit & & \vtop{\hbox{\strut Youtube's next video}\hbox{\strut recommendation}} \\[6ex] \hline

Hybrid & \vtop{\hbox{\strut Google Scholar search}\hbox{\strut Semantic Scholar search}\hbox{\strut Gomez-Uribe and Hunt 2016 \cite{gomez_hunt_2016}}\hbox{\strut Hersh at al. 2001 \cite{hersh_2001}}}
& \vtop{\hbox{\strut Facebook's user's feed}\hbox{\strut Gama at al. 2014 \cite{gama_2014}}} \\ [6ex] \hline
\end{tabular}
\caption{Popular search and recommendation systems divided to explicit / implicit / both explicit and implicit (hybrid) and one-time or lifetime querying.}
\end{center}
\end{table}

\section{Project-Based Search}
Project-based search methodology addresses concept drift in information systems. It has two requirements: First, the user must provide explicit information as to the context for the query. Second, the user has multiple queries.

Specifically, in searching for research papers, both criteria are met. Researchers work on research projects in parallel, and these serve as contexts for their searches. They almost invariably continue to search for relevant research literature for at least the lifetime of a project. When examining a sequence of queries without knowledge of the research topics, the queries seem to be very varied. But given explicit knowledge of the search intent (the concept), the variance is reduced. In this case, project-based search becomes handy.

The project-based search methodology is based on two main components: a one-time search and a personalization algorithm. Rivendell is a meta-search engine designed for searching for research papers implementing the project-based search methodology and the required algorithms the methodology is based upon. In addition, a new query suggestion in the scope of a query is implemented to further utilize the information the user explicitly provided for the personalization algorithm. 

To examine the influence of project-based search on information system (search and recommendation systems), the algorithms used in Rivendell are relatively simple comparing to the state of the art algorithms used in natural language processing (NLP) and information retrieval (IR). This allows investigating the direct influence of project-based search which out taking into consideration the complex dynamics that emerge by using advanced machine learning methods.

\subsection{One-time Search Algorithm}

\begin{figure}[p]
\centering
\includegraphics[width=1.0\textwidth]{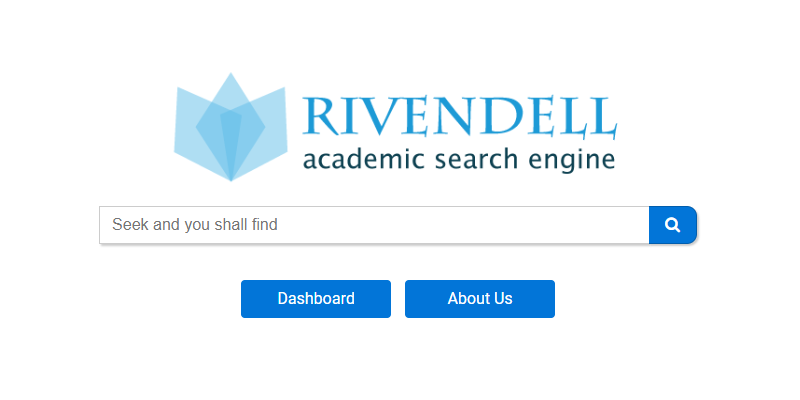}
\caption{The user interface of a one-time search in Rivendell.}
\label{fig:one_time_search_ui}
\end{figure}

A one-time search, which does not use personalization at all, provides results as do other search engines. Figure \ref{fig:one_time_search_ui} shows the user interface of the one-time search page at Rivendell.

Rivendell takes advantage of PubMed and arXiv using their APIs. The process takes advantage of the inner search algorithms of each service and then adds them a Boolean search logic with additional filtering and reorder done by personalization technique driven by the historical queries and labeling of the user. The system allows the user to search using a query language based on ``Boolean searching" \cite{JoeranBeel}. We used the Boolean searching method as it will be exploit in the next query suggestion algorithm (see Section 4.3) and as it shown to be an effective method \cite{JoeranBeel}.
When a user searches in Rivendell, the query can include ``and", ``or", ``and not". The system does not allow ``or not" or ``not" operators, as they are meaningless in the search operation. For example, for the query: \(``A \vee (B \wedge C) \wedge \neg E"\), the system will return: \(``A \vee B \wedge A \vee C \wedge \neg E"\), as it is not allowed to request all the results expect the ones contains \(E\) as reflected from \(\wedge \neg E"\)). After retrieving the expended format of the query, Rivendell separates each sub-query between two  logical ``and" operators. In case of a query without an ``and" logical operator the sub-query is the whole query. Following our example, the query will be  \(``\{A \vee B, A \vee C, \neg E\}"\). Each similar element is then separated again according to the ``or" logical operator to ``atomic terms". For example  \(``\{\{A, B\}, \{A, C\}, \{\neg E\}\}"\). Finally, all the atomic terms are divided into two groups: one for atomic searches with ``not" logical operator operating on them and one for the rest. For example: \(``\{\{\{A, B\}, \emptyset \} , \{\{A, C\},  \emptyset\},  \{ \emptyset, \{\neg E\}\}\}"\).

Using two API calls for each database, Rivendell selects all the abstracts containing all atomic terms. After receiving all the lists of abstract related to each atomic term, it then merges the list of results for each ``or" operator. Given the lists of each sub-query contains only ``or" operations, Rivendell intersectes them all; if the list belongs to the ``not" group, then the intersection between both lists \((A; B)\) returns only \((A \cup B) / (A \cap B)\) Otherwise, just \((A \cap B)\). The result of this process is a list of abstracts, which is passed to the personalization stage, which takes into account all the previous interactions in the same project.

\subsection{Personalization in Rivendell}

\begin{figure}[p]
\centering
\includegraphics[width=1.0\textwidth]{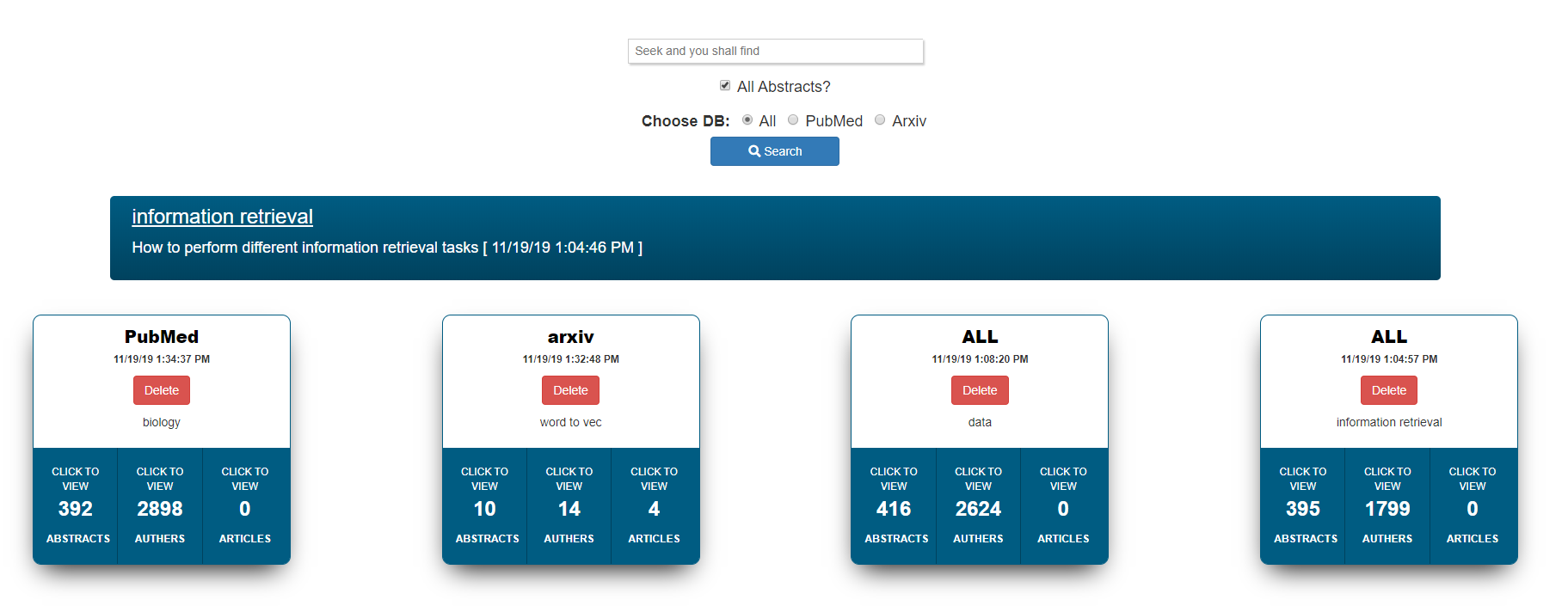}
\caption{The user interface of a single project with the project-based search panel.}
\label{fig:single_project_view}
\end{figure}

\begin{figure}[p]
\centering
\includegraphics[width=1.0\textwidth]{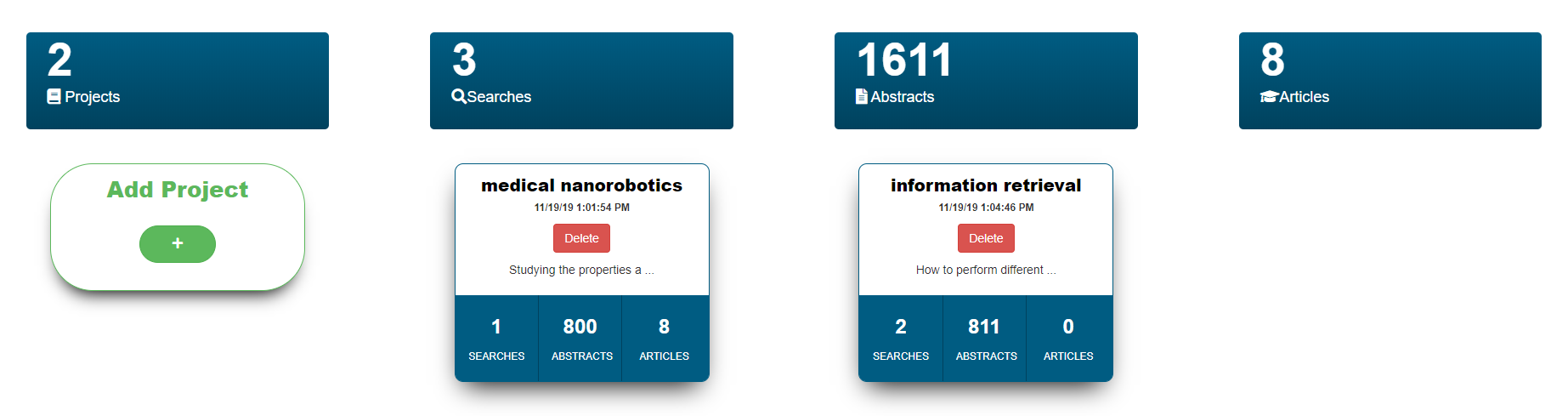}
\caption{Rivendell's listing all of a user's project.}
\label{fig:project_summery_ui}
\end{figure}

Users with a topic in mind take advantage of Rivendell's project-based search which learns their preferences while searching. By opening a ``project", the user explicitly declares to the system that all the queries inside this project should be learned together. Figure \ref{fig:project_summery_ui} shows the user interface listing all the user's projects. Figure \ref{fig:single_project_view} shows the user interface of a single project, and the project-based search panel of this project. Older projects with some statistics about them can be found and reviewed inside the project page. This allows the user to pick its scope and help the system to learn each research topic separately, allowing to look for different research topics simultaneity in different projects without introducing noise to each research topic learning process. This time, the labeling of relevant and irrelevant results effects both the results order in this query and later the filtering and order of future queries' results.

The personalization algorithm is used after the Boolean search and the API calls to the inner search engines. The query results (a set of abstracts $A$) are ranked and filtered by the personalization algorithm. The algorithm assumes access to all the previous queries $q_1,...,q_k$ in the same project, the returned document sets for these queries, and whether the user marked each abstract as relevant or irrelevant. We mark the set of relevant abstract for query $q_i$ as $A^p_i$, and the irrelevant ones as $A^n_i$. Each abstract $a\in A$ is represented as a term vector \cite{bow}, where each term is weighted by its frequency (stop-words from a predefined list of function words and common words are discarded).

We then compute a score for each returned abstract $a\in A$ by taking into account the query similarity to the previous queries, as well as the average document similarity to the relevant documents of that query:

\[ score(a) = \sum_i sim^q(q,q_i) \frac{1}{|A^p_i|}\sum_{a' \in A^p_i} sim^d(a,a') \]

The document similarity $sim^d$ is the cosine distance between the term vectors, and the query similarity $sim^q$ is the Monge Elkan algorithm~\cite{MongeElkan} with Levenshtein distance~\cite{LevenshteinVladimir} (the queries' words are stemmed, and a logical ``not'' operator in one of the queries---but not the other---negates the returned score) \cite{query_formula, query_context_1, atypical_search}.

We then compute the mean and standard deviation of the scores for all documents $a \in A$, and discard documents whose scores are more than \(x\) standard deviations below the mean (we currently use 2, chosen empirically). This filtering stage can be too aggressive when the project's history information is sparse or non-existent. We therefore apply this filter only if it retains at least a fixed percentage of the results in the list (we currently use 60\%, chosen empirically). 

\subsection{Next Query Suggestion}

\begin{figure}[p]
\centering
\includegraphics[width=1.0\textwidth]{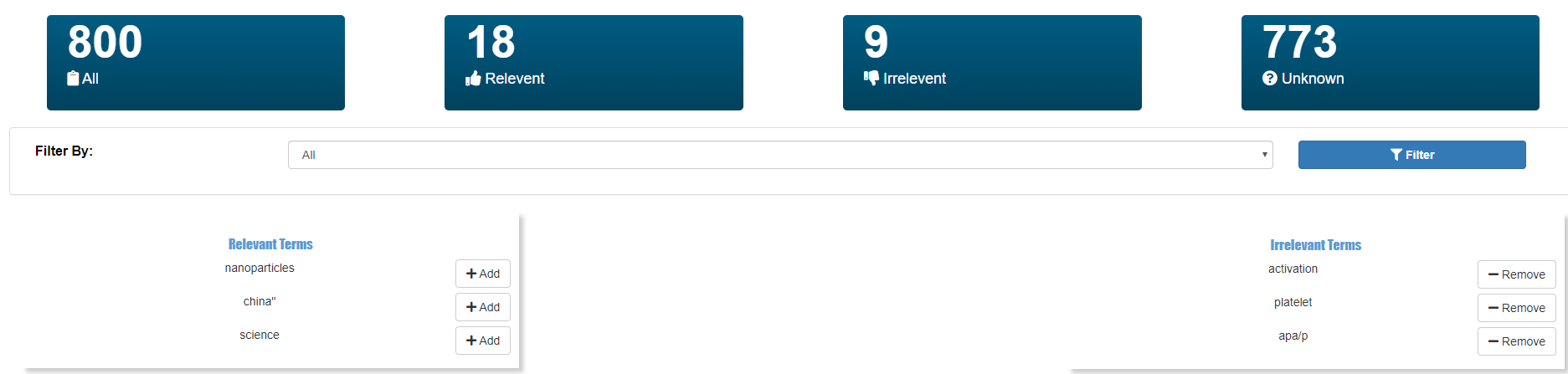}
\caption{The user interface of the suggested terms to be added (on the left side) or removed (on the right side) from the query to improve the relevancy of the results according to the user's binary tagging of the query's results. }
\label{fig:suggestion_panel}
\end{figure}

In the context of a query inside a project, the user can re-order the results according to labeled results in the same query. In addition, they get quick access to all the relevant papers in the account. These features encourage the user to label results. The system uses these labels in each query to find and suggest common terms by splitting each results into a set of terms, scoring each term according the the result's label and finally sort the terms according to there scores. Adding or removing the suggested terms from the query should provide results with higher precision in the following query, as shown in figure \ref{fig:suggestion_panel}.

The suggestion algorithm is designed to improve the converging time of a user's results by suggesting the next query to run according to the results of the last query. The suggested query is in the form of the original query with an addition or deletion of a term from the query by using the Boolean search mechanism in the search algorithm. The suggestion system is symmetrical to relevant and irrelevant abstracts and is handled identically. The relevant data set provides the additional suggestions the system provides, while the irrelevant data set provides the deletion suggestions.

For example, if a user query is ``medical nanorobotics and gold" and the system decides on a new term ``X," then it generates the query: ``medical nanorobotics and gold \textbf{and X}" or on the second option: ``medical nanorobotics and gold \textbf{and not X}".

The term ``X" is found in the following way: First, the algorithm provides the relevant abstracts from a given query $A^p_q$. Each abstract $a \in A^p_q$ is represented as a term vector, where each term is weighted by its frequency \cite{bow} (stop-words from a predefined list of function words and common words are discarded). Then, a joined vector $Ta$ of all the vectors $a \in A^p_q$ is calculated. The terms in vector $Ta$ are ordered by there frequency. At this point, the mean \( E[Ta]\) and standard deviation \(SD[Ta]\) of the terms' frequency is calculated. We filter the words $w \in Ta$ with a score lower than one z-score (\(\frac{w - E[Ta]}{SD[Ta]}\)). If any words have passed the threshold, they become the suggestions for the next query.

\subsection{Rivendell} 

Rivendell\footnote{http://rivendellscholar.info} is a meta-search engine searches within arXiv and PubMed using the project-base search approach. it has been designed to provide a better solution for the searching process of academic papers considering the unique characteristic of the searching process of researchers: a quick review of a research topic (exploration), finding specific results addressing a narrow topic in question, as the goal of the user change over time because of the nature of the research process. Rivendell's main feature is the ability to query academic databases when providing a context (project) with the query to run on-line individual personalization algorithms which filter and reorder the original results from the other academic databases it based on using the user's usage information. Rivendell addresses concept drift by using one-time search and project-based search.

Rivendell provides both one-time and project-based search in the form of ``quick" and ``project related" search options, respectively. Quick search is more suitable for exploration of a topic without the need to provide additional explicit data. Project related search is useful when the user has already opened a project in the system and wishes to query something inside the scope of this project. The user picks which project the query is related to and the query itself; this additional information allow the system use project-based searching.

\begin{figure}[p]
\centering
\includegraphics[width=1\textwidth]{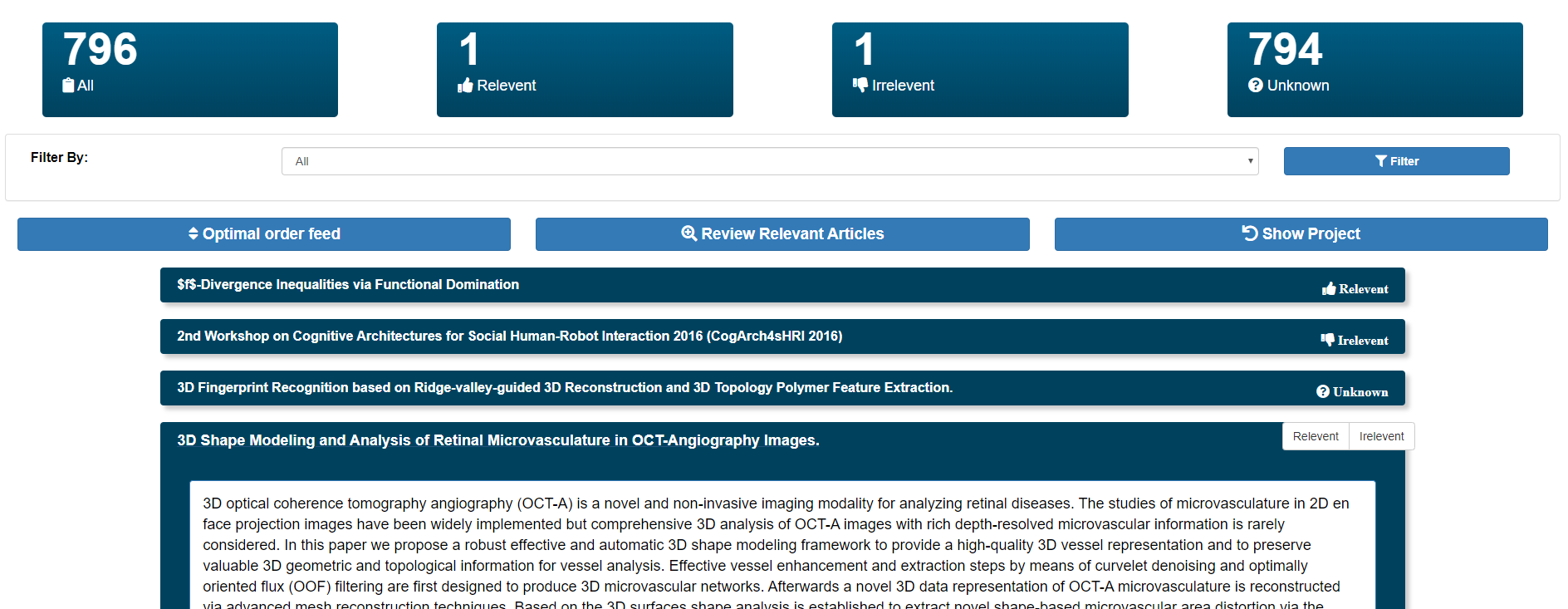}
\caption{The user interface of a one-time search's result in Rivendell.}
\label{fig:one_time_search_results_ui}
\end{figure}

Rivendell allows the user to label relevant and irrelevant results, as shown in figure \ref{fig:one_time_search_results_ui}. Here, we use a query personalization scope labeling which affects only the specific query by reordering unlabeled results similar to the results found relevant and differ from the results that found irrelevant; allowing the user to explore more similar or diverse results without adding bias to further exploration queries implementing the next query technique (described above).

All three algorithms used in Rivendell are well-known and broadly used. These algorithms are not state of the art in natural language processing or information retrieval. These algorithms chosen for this version of Rivendell due to their simplicity, which allows to more easily investigate the influence of the project-based search mechanism in the system level.

\section{Experiments and Results}

\begin{figure}[htbp]
\centering
\includegraphics[width=6cm]{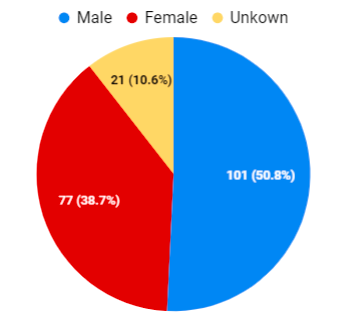}
\caption{Distribution of participants' gender}
\label{fig:gender_destrebution}
\end{figure}

\begin{figure}[p]
\centering
\includegraphics[width=1.0\textwidth]{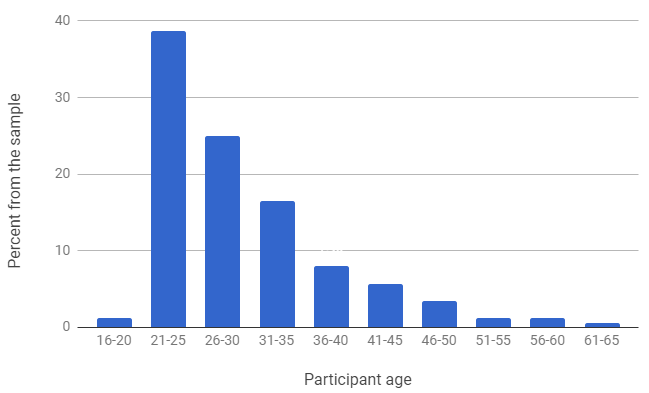}
\caption{Distribution of participants' age}
\label{fig:participants_ageg}
\end{figure}

The evaluation was conducted on Rivendell's users in the live (production) system. The results of each experiment were saved in the inner experiment sub system in Rivendell as logs of users’ actions [entering a page, running a search, marking a result]. The experiments were done without supervision to ensure the result are bias-free. The users taking part in the experiment were analyzed based on gender, age, and institute. Overall, there were 199 participants (60 in the first experiment, 100 in the second experiment, and 39 in the third experiment). The participants did not get paid for their work. Figure \ref{fig:gender_destrebution} shows the distribution of the participants' gender, with more male than female.
Figure \ref{fig:participants_ageg} shows the distribution of the participants' age, most of them were between 21 to 25 years old. The mean age was 30.5 (sd. 8.52).

\begin{figure}[p]
\centering
\includegraphics[width=1.0\textwidth]{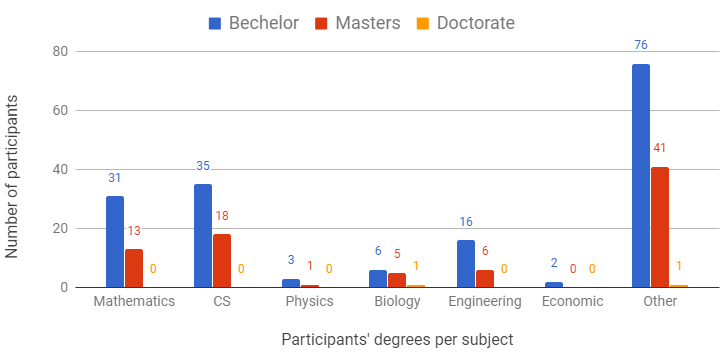}
\caption{Distribution of academic background of experiment participants. }
\label{fig:participants_academic_bg}
\end{figure}

To be able to successfully participate in the experiments it is assumed that the users should have an academic background to be able to read academic papers and evaluate their relevancy to the query. We, therefore,
required that participants had at least a bachelor degree. The participants declared their academic background.
Overall, we had 85 participants with an only bachelor degree, 82 with a masters degree, and 2 with a doctorate degree. 95 ( 48\%) of the participants studied in Israeli universities, 56 (28\%) at Russian universities, the other 48 (24\%) either did not enter a legitimate university name or entered a current work position.
Figure \ref{fig:participants_academic_bg} shows the distribution of the academic background (degree and field of study) of the participants.
The institutes with the biggest number of participants are: Moscow University with 57 (28.5\%) participants, Bar Ilan University with 51 (25.5\%) participants, Tel-Aviv University with 27 (13.5\%) participants, and Open University of Israel with 13 (6.5\%) participants. All other institutes with less than 4\% each.

We used Facebook and Linkedin to recruit participants for the experiments. On both social sites, we published posts in Hebrew and English, asking people with an academic background to participate in our experiment. Furthermore, posts in professional groups like "machine and deep learning Israel" on Facebook and "Algorithms and Optimization" on Linkedin were published all in the same day to attract as much attention from people who are members of multiple connected professional groups. Finally, personal messages were sent to individuals on Linkedin who declared their academic background in their profile pages to be Bachelor’s degree or higher and were connected with at least one of the authors, asking them to partake in our experiments.

In all the experiments, we started by presenting the participants with a page with a short paragraph describing the research and its purpose. After reviewing the terms and conditions, the users then agreed to them and processed to a registration page; The data of this page used for the statistical analysis of the participants presented above. Each participant is randomly assigned to one of the three experiments. Before the experiment itself, a page with a description of the task ahead was presented. The user had to the task and only by confirming that it read it, the experiment starts.

It is possible to separate the system into eight features that can be controlled while using the searching mechanism: subject of search, learning set, whether the ML method used, which database is queried, how many results asked to be return, are Boolean querying language is used, the query itself, and which user interface shows the results. At the beginning of each experiment, we declare which features are used and how.

\subsection{Baseline}

We wish to compare the performance of \textit{Project based} search in \textit{one-time} and \textit{life-time} searches. First, we establish a baseline between the performance of the one-time search without the personalization nor the project-based search comparing to arXiv and PubMed. This experiment tested whether Rivendell, as a meta-search engine, provides results at least as good as PubMed and arXiv that it depends on. We wished to compare to Google Scholar and Semantic Scholar as well but neither offers an API, and their terms and conditions explicitly forbid meta-searches using their services. One may consider an experiment in which a user querying one of these search engines while using the same query in Rivendell to compare the results. This kind of experiment does not provide a valid comparison of the search engines performance \cite{ui_www, ui_trust}, as a result of the different user interface, experiment flow, and the data presented for each result. Such kind of experiment will evaluate the overall performance of the systems tested rather than the differing in the searching methods these systems use.

We recruited 60 users for this experiment. 35 from an exact science background, 21 from biological oriented science and, four in other fields of science. The 60 users were asked to enter a query in the fields of exact or natural science but no check has been done on the inputs. Then, the system carried out three independent searches: the first two, using PubMed and arXiv APIs; we provided the query from the user to the API and just converted the result to a title and abstract. The third search was done using Rivendell’s searching algorithm when the personalization and boolean querying are disabled; the search used both PubMed and arXiv.
The measurement of how many results are relevant was done by asking the user to mark for each query, a pseudo-randomly picked 10-result list, and compare the percent of relevant results across the different systems.

All the outcomes were collected to a single list and were presented to the user in random order on a blank page outside Rivendell with minimal design to neglect as much as possible the effect of user interface on the results. The user had to mark the first ten outcomes as relevant or irrelevant. Then, the user repeated the process at least two more times of searching a new query and tagging the first ten outcomes. Then, they were allowed to continue as long as they wish.

All these users had executed three searches and marked, at least, thirty abstracts.
Initially, PubMed searches showed worse results relative to arXiv and Rivendell. We believe that this is because users without a biology background failed to evaluate the relevance of PubMed papers. We, therefore, normalized the results based on the participants' background as discussed below, shown in Table~{tab:q1}.
Both normalized and unnormalized results show that Rivendell's results are just as relevant as those from its component search engines (i.e., no noise is being introduced.)

\begin{table}
    \centering
    \begin{tabular}{|l|c|c|c|}
    \hline
	Search Engine & PubMed & arXiv & \bf{Rivendell} \\
	\hline
	precision (\%) & 32.45 & 46.15 & 47.76 \\
	\hline
	Normalized precision (\%) & 45.43 & 46.15 & 47.76 \\
	\hline
    \end{tabular}
	\caption{Mean percentage of query's precision. Rivendell's one-time search is as good (indeed, marginally better) as that of PubMed and arXiv.}
	\label{tab:q1}
\end{table}

In Table~\ref{tab:q1}, PubMed shows worse results relative to arXiv and Rivendell. We argue that the ability of users without a background in biology to properly evaluate the relevance of results from PubMed is questionable. Of the 60 users, 35 searched for research topics from an exact science and had the proper background to properly evaluate the relevance of the results. On the other hand, the other 25 users did not necessarily search for biological research topics but on a variety of research topics. Therefore, there is a bias in natural science queries. To properly evaluate the results, one needs to balance the bias in unrelated backgrounds. This can be done by a normalization of the results according to the participants' background. We normalize all the columns according to the following formula:

\[ QP' = \frac{\max\limits_{RSB}({PubMed, arXiv, Rivendell}) \cdot QP}{RSB} \]
When \(\max\limits_{RSB}({PubMed, arXiv, Rivendell})\) stands for the number of maximum subjects with a background allowing to properly evaluate the relevancy of a result according to a given query in all the search engines, \(QP\) stands for the query's precision of quires in a given search engine, \(RSB\) stands for the number of subjects with the background allowing to properly evaluate the relevancy of a result according to a given query, and \(QP'\) stands for the normalized query's precision of quires in a given search engine, respectively. The results of the experiment after normalization was presented in Table~\ref{tab:q1}. A two-tailed t-test with \(\alpha=0.05\) has been conducted and shows no statistical significance between the results of Rivendell and arXiv or Rivendell and PubMed. One can conclude that Rivendell's results in these settings are similar to both arXiv and PubMed. 

\subsection{Project Based Search vs Life Time Search}

Given the personalization algorithm presented earlier, we compare project-based search to life-time search. This experiment is testing the hypothesis that project-based search provides better results than life-time search using Rivendell’s personalization algorithm as presented earlier. At the beginning of the experiment, each one of 100 users was selected for one of four groups randomly (each group with 25 users). A ``base" group in which does not use the personalization algorithm and operates as the baseline group. A ``random" group in which provided by a random sample of the previous queries of the user. The sample size in the second and third searches is a single query, on the fourth, fifth, and sixth searches, the sample set is two queries. A ``project" group in which provided by previous queries by context. Meaning, the first three queries together and the last three queries together. A ``lifetime" group in which provided by all the previous queries of the user.

The users were asked to enter six queries---three and three---in the fields of exact or natural science (no check has been done on the input). The first three on a research topic they had in mind and the following three searches on a different research topic in the same field. After each query, the users were asked to mark ten results as relevant or irrelevant. After reviewing the last query, the users finished the experiment by clicking on the ``end experiment" button. The results from each query presented to the user in a random order to ensure equally distributed sampling of the results from the set of possible results the searching algorithm provides. We filter outliers from each group to remove users that have done the experiment unfairly tagging all the results as relevant or irrelevant. Overall, We have removed two users from the data set, one from the ``project" group and one from the ``random" group.

\begin{figure}[p]
	\centering
	\includegraphics[width=1.0\textwidth]{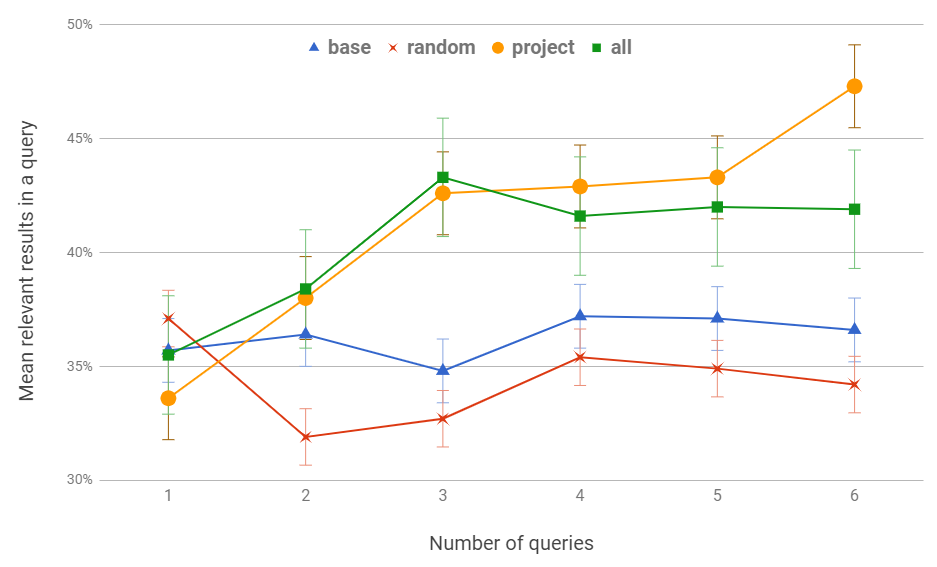}
	\caption{Relevant results vs number of queries. Error bars indicate the standard error. Switching from the first topic (queries 1--3), to the second (queries 4--6) project based search improves, while life-search declines.}
	\label{fig:q2}
\end{figure}

Figure~\ref{fig:q2} shows the mean percentage of relevant results for each group, versus the number of queries. The x-axis measures the number of queries from the beginning of the experiment and the y-axis measures the mean precision of relevant results (percentage). Each point represents the mean precision of relevant results over 25 samples.

It is possible to see that the baseline and the random groups are similar in behavior but the random group is constantly worse than the baseline. Such phenomena happen from the random noise inserted to the personalization algorithm which chooses irrelevant results in later unrelated queries. The results from the random group show that meaningful choice is required for a proper personalization algorithm.  A one-tailed t-test with \(\alpha=0.05\) has been conducted and shows a statistical significance between the results of the project group and both base \(p=0.0311\) and random groups \(p=0.0134\). 

The first three queries are on the same topic. One can observe an almost identical behavior between the ``project" and the ``life-time" group from the first to the third query; in both cases, the algorithm works identically. However,
when switching to the second topic in query 4, there is a change in behavior between the two groups and indeed the ``project'' group improves while the ``life-time" group does not. This is caused by the changes in topic (context) between the two sets of queries. The personalization algorithm can take this context switch into account when using a project-based search. By the sixth query, the difference is 12.89 percent between the two groups. One can see that explicit context switching of the user clearly improves the personal results of the search engine with the personalization algorithm.

One may notice that there is no decrease in precision of relevant results between the third and fourth query in the ``project" group as expected when starting a new project as accrue at the ``life-time" group. A paired t-test between ``project" group's third query and the fourth result shows there is no statistically significant changes \((p=0.971)\). Repeating for the ``life-time" group provides similar results \((p=0.867)\). Therefore, the local change of a single query between two subjects is not significant enough. 

\subsection{Next Query suggestion }
To reduce the number of needed queries to receive a given precision, we evaluate Rivendell's query suggestion component. At the beginning of the experiment, 30 users were assigned to three groups: A \textit{search only} group in which cannot use nor see the suggestions from the system and need to write a free-text query; this group operates as the baseline. A \textit{suggestion only} in which a user firsts writes a free-text query and then each following one has to be picked from the suggested options by the system. A \textit{suggestion and search} in which present to the users the suggestion from the system and allows the user to pick one of the suggested queries or to write a free-text query.

The users were asked to enter a query in the fields of exact or natural science but no check has been done on the inputs. At the experiment, we asked users to enter at least five queries. After each query, the users are asked to mark ten (or more) results as relevant or irrelevant by clicking on the proper button. After reviewing the results of each query, the users had to operate according to the group they were assigned to. After reviewing the last query, the users finished the experiment by clicking on the ``end experiment" button. The results on the page were ordered randomly to ensure equally distributed sampling of the results from the set of possible results the search algorithm provides.

\begin{figure}[p]
	\centering
	\includegraphics[width=1.0\textwidth]{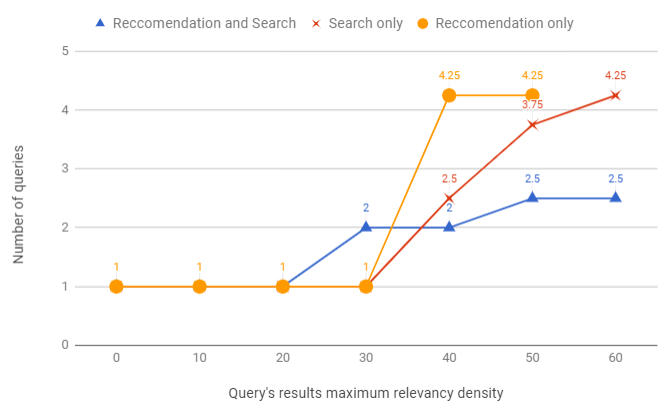}
	\caption{Minimal query for a given relevancy precision}
	\label{fig:q3}
\end{figure}

Figure~\ref{fig:q3} presents how fast an average user converged to the relevant results. Each point represents the mean precision of relevant results over 10 samples (precision at 10).
 
The ``search only" group is the baseline group, as this is naive, without a suggestion case, and shows the learning curve of users of the system’s search engine. The ``suggestion only" group checks the ability of the system to correctly find the next query the user wishes to run to improve the results of the system with the same question in mind. The ``suggestion and search" group is representative of the actual case in Rivendell.

We note that there is up to a thirty percent precision in all groups combined, because they indeed use a free text query for the first run and as we saw in the ``Project-based search" experiment on the first query, the users achieve around 35 percent relevance precision. From this point, the groups separate the ``only search" group shows a learning curve which converges over time. When looking at the ``suggestion only" group, there is a sharp increment from one to five queries when crossing the thirty percent precision. Such sharp change indicates that the system {poorly} learns by itself the next needed query for the user.

The ``suggestion and search" group shows a learning curve much better than the ``search only" group, and the ``suggestion only" groups. Surprisingly, we found that only eight percent of the queries of this group were generated using the suggestion system which raises questions as to what caused the users in this group to improve in comparison to the ``search only" group.

We, therefore, conducted a final small experiment, with nine additional users.  They started the experience as usual but this time, after they review the results and mark them, we asked them to write the next query they wish to search. After a user writes the next query, the query suggested by Rivendell was shown, and the user was asked again to write the next query.  In effect, we tested whether users rethink their queries after being shown Rivendell's suggestion.

The results from this experiment explain why the combined ``suggestion and search" group improved on both of its component parts. It turns out that users changed their queries after seeing the suggestion of Rivendell (close to 80\% of the time), even though they did not blindly follow Rivendell's suggestion. This resulted in the synergistic effect of combining Rivendell's suggestion and the users'. The results then showed a thirty percent learning curve improvement in the user’s ability to converge to a majority of relevant results from the results the system provides: two and a half queries on average for the "suggestion and search" vs three point twenty fine for the baseline.

\section{Summary}
This paper presented Rivendell, a novel meta-search engine specializing in finding scientific papers. Rivendell explores the use of project-based search, where the user provides explicit context (the research project), and thus assists the search in overcoming concept-drift in multiple queries. In addition, Rivendell is able to generate suggestions for new queries, based on explicit feedback from the user as to the relevancy of previous queries' results.

We show that in a cold-start setup, project-based search outperforms life-time search, and causes users to rethink their intended future queries, resulting in much faster improvements to the precision of relevant results in the output set. Rivendell is currently in use by a few hundred users a month.

This study examines the influence of project-based search using explicit tagging by users to avoid concept drift and improvement of quires' precision in information systems (search and recommendation). The natural language processing (NLP) and information retrieval (IR) algorithms used in the study may be considered relatively simple compared to the state of the art \cite{gpt}. For example, in the one-time search algorithm, we used the Boolean search method (see Section 4.1) while other methods such as \textit{ranked retrieval} may be considered more effective \cite{ranked_retrieval}. In future research, we will investigate the performance of Rivendell after replacing the current NLP and IR algorithm with state of the art algorithm. In addition, community learning using shared groups with multiple tagging of each query's results is another promising direction.  

\section{Funding Sources}
We gratefully acknowledge partial funding from ISF group \#2306/18.

\printbibliography

\end{document}